\documentstyle[aps,prl,twocolumn,floats,graphicx]{revtex}

\begin{document}
\draft
\preprint{HEP/123-qed}
\wideabs{
\title{Low Temperature AC Conductivity of Bi$_2$Sr$_2$CaCu$_2$O$_{8+\delta}$}
\author{J. Corson, J. Orenstein}
\address{Materials Sciences Division, Lawrence Berkeley National Laboratory and Physics Department, University
of California, Berkeley, California 94720}
\author{J.N. Eckstein}
\address{Department of Physics, University of Illinois, Urbana, Illinois 61801}
\author{I. Bozovic}
\address{Oxxel GMBH, D-28359 Bremen, Germany}
\date{\today}
\maketitle
\begin{abstract}
	We report measurements of anamolously large dissipative conductivities, $\sigma_1$, in Bi$_2$Sr$_2$CaCu$_2$O$_{8+\delta}$ at low temperatures.  We have measured the complex conductivity of Bi$_2$Sr$_2$CaCu$_2$O$_{8+\delta}$ thin films at 100-600 GHz as a function of doping from the underdoped to the overdoped state.  At low temperatures there exists a residual $\sigma_1$ which scales with the $T=0$ superfluid density as the doping is varied.  This residual $\sigma_1$ is larger than the possible contribution to $\sigma_1$ from a thermal population of quasiparticles (QP) at the d-wave gap nodes. 
\end{abstract}
}

The conductivity of a superconductor combines two components: the superfluid and the quasiparticle (QP) excitations.  As one cools a superconductor from $T_C$ towards $T=0$, the fraction of the charge carriers excited out of the superfluid decreases towards zero.  Therefore, at low temperatures the finite frequency conduction is increasingly dominated by the response of the superfluid rather that the dissipative QP conductivity.  This is exemplified in YBa$_2$Cu$_3$O$_7$ (YBCO) by the low temperature behavior of $\sigma_1$\cite{Hosseini}.  Hosseini, {\em et al.} find $\sigma_1(T)$ to exhibit a broad peak below $T_C$ and approach zero $\sim$ linearly in the $T=0$ limit.  The peak in $\sigma_1(T)$ is thought to result from a competition between the increasing QP lifetime, $\tau_{QP}$, and the decreasing QP density, $n_{QP}$, as $T\rightarrow 0$.  

In the cuprate Bi$_2$Sr$_2$CaCu$_2$O$_{8+\delta}$ (BSCCO), however, very different low temperature behavior has been observed.  $\sigma_1$ of optimally doped BSCCO single crystals at 35 GHz approaches a constant value which is $\sim 10$ times $\sigma_{1 - Normal}$ at the lowest temperatures measured ($\sim$~5K)\cite{SFLee}.  Using Time-Domain-THz-Spectroscopy we extended measurements of the complex conductivity, $\sigma=\sigma_{1}+i\sigma_2$, in BSCCO to higher frequencies and different carrier concentrations.  We studied a set of BSCCO thin films varying from under ($T_{C}=33K$) to over-doped($T_{C}=74K$).  Figure 1 is a semi-log plot of $\sigma_1$(140 GHz) versus $T$.  $\sigma_1$ is seen to have two components: a peak centered near $T_{C}$ superimposed on a broad background.  This peak, less prominent as doping increases, results from thermally generated phase fluctuations\cite{Corson}.  The background $\sigma_1$ is seen, at all dopings, to increase to a value well above $\sigma_{1 - Normal}$ and does not decrease to zero as $T\rightarrow 0$.

\begin{figure}[h]
\begin{center}
\leavevmode
\hskip -.3in
\includegraphics[width=1.1\linewidth]{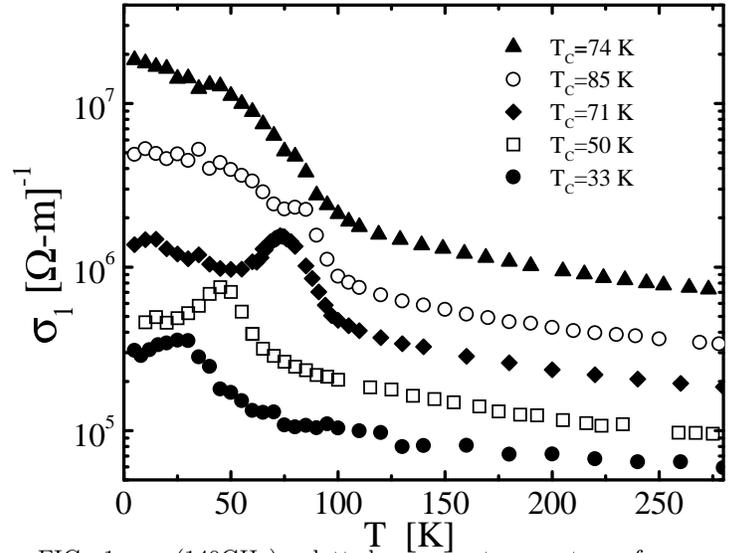}
\vskip -.05in
\caption{$\sigma_1$(140GHz) plotted versus temperature for BSCCO films from underdoped $T_{C}=33K$ to overdoped $T_{C}=74K$.
}\label{figurename}\end{center}\end{figure}

In the optimal and overdoped samples the superfluid density, $\rho_S$, is seen to decrease linearly with $T$\cite{Future}.  If we ascribe this low $T$ decrease in superfluid density $\Delta\rho_{S}\sim T$ with a $n_{QP}=n_{Total}(T/T_{C})$ \cite{Hosseini,SFLee}, then the fraction of $\sigma_1$ which can be attributed to the QP, $\sigma_{1-QP}$, must tend to zero as $T\rightarrow 0$.  We illustrate this using the Drude form for the QP conductivity, $\sigma_{1-QP}=n_{QP}\tau_{QP}/(1+(\omega\tau_{QP})^{2})$, with $n_{QP}=n_{Total}(T/T_{C})$.  Using our value of $\sigma_{1-Normal}$ and the measured normal state value of $\tau_{QP}$ \cite{Tanner} we can determine $n_{Total}$.  Following the recent ARPES measurements by Valla {\em et al.} we assume the $T$ dependence of $\tau_{QP}$ to be $1/ \tau_{QP}(T) \sim T+T_{0}$ \cite{Wells}. 

Figure 2 shows, for the $T_{C}=85K$ sample at 200 GHz, $\sigma_{1-QP}$ from this model, $\sigma_{1-Total}$, and the residual conductivity not due to the QP, $\sigma_{1-Res}\equiv\sigma_{1-Total}-\sigma_{1-QP}$.  $\sigma_{1-QP}$ accounts for all of $\sigma_1$ down to 100 K, the temperature identified in previous work with the onset of identifiably non-zero bare $\rho_{S}$ \cite{Corson}.  $\sigma_{1-Res}$ appears at 100 K, exhibits the thermal phase fluctuation peak at 80 K and then rises linearly in $T$ over the rest of the temperature range down to zero.  

Finally, we find $\sigma_{1-Res}$ to be proportional to $\rho_{S}(T=0)$ over our range of doping from the overdoped to the $T_{C}=50K$ underdoped sample. The width of $\sigma_{1-Res}(\omega)$ in frequency seems to be $\le$ 200 GHz for all of the samples studied.

If, instead of the Valla {\em et al.} $\tau_{QP}(T)$, we use a more conventional form $\tau_{QP}(T)\sim T^{\alpha}+T_{0}$, $\sigma_{1-QP}$ develops a peak below $T_{C}$ and $\sigma_{1-Res}(T)$ is no longer linear in T from 80K to 5K.  However, $\tau_{QP}$ still approaches a small constant as $T\rightarrow 0$ and the other result is unchanged:  $\sigma_{1-QP}$ is still a small fraction of $\sigma_{1-Total}$ at low temperatures.

\begin{figure}[h]
\begin{center}
\leavevmode
\includegraphics[width=\linewidth]{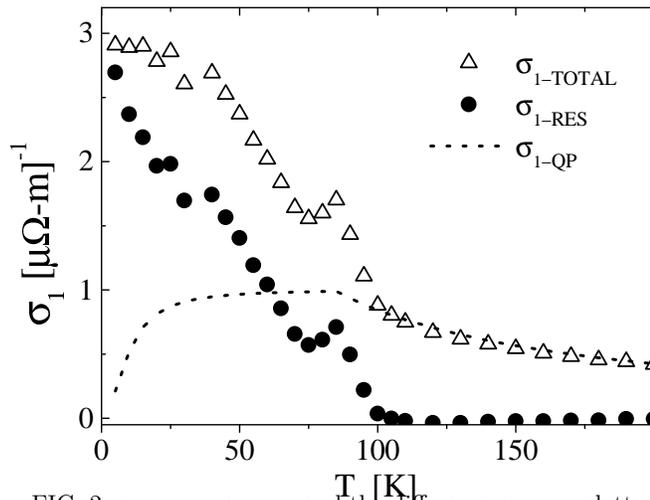}
\vskip -.05in
\caption{$\sigma_{1-Total}$, $\sigma_{1-QP}$ and the difference $\sigma_{1-Res}$ plotted versus temperature for the $T_{C}=85K$ sample at 200 GHz.
}\label{figurename}\end{center}\end{figure}

Since $\sigma_{1}$ is too large to be ascribed entirely to the QP regardless of the choice for $\tau_{QP}$, and $\sigma_{1-Res}\sim\rho_{S}(T=0)$ as one might expect for a collective excitation in the superfluid, we speculate that this low $\omega$, low $T$ peak in $\sigma_1$ could signal observable quantum fluctuations in the BSCCO system.

\acknowledgments
This work was supported by NSF, DOE and ONR


\end{document}